\documentclass{ws-ijqi}

\newcommand{\ket}[1]{| #1 \rangle}

\newcommand{\pra}{{\it Phys. Rev. A~}}

\newcommand{\prl}{{\it Phys. Rev. Lett.~}}

\newcounter{myctr}
\def\myitem{\refstepcounter{myctr}\bibfont\noindent\ifnum\themyctr>9\else\phantom{0}\fi\hangindent17pt\themyctr.\enskip}

\begin{document}

\markboth{R.~Ionicioiu and W.J.~Munro}
{Constructing 2D and 3D cluster states with photonic modules}

\catchline{}{}{}{}{}

\title{CONSTRUCTING 2D and 3D CLUSTER STATES WITH PHOTONIC MODULES}

\author{RADU IONICIOIU}

\address{Hewlett-Packard Laboratories, Long Down Avenue\\
Stoke Gifford, Bristol BS34 8QZ, UK\\
radu.ionicioiu@hp.com}

\author{WILLIAM J. MUNRO}

\address{Hewlett-Packard Laboratories, Long Down Avenue\\
Stoke Gifford, Bristol BS34 8QZ, UK\\
bill.munro@hp.com}
\address{National Institute of Informatics, 2-1-2 Hitotsubashi\\
Chiyoda-ku, Tokyo 101-8430, Japan}

\maketitle

\begin{history}
\received{Day Month Year}
\revised{Day Month Year}
\end{history}

\begin{abstract}
Large scale quantum information processing (QIP) and distributed quantum computation require the ability to perform entangling operations on a large number of qubits. We describe a new photonic module which prepares, deterministically, photonic cluster states using an atom in a cavity as an ancilla. Based on this module we design a network for constructing 2D cluster states and then we extend the architecture to 3D topological cluster states. Advantages of our design include a passive switching mechanism and the possibility of using global control pulses for the atoms in the cavity. The architecture described here is well suited for integrated photonic circuits on a chip and could be used as a basis of a future quantum optical processor or in a quantum repeater node.
\end{abstract}


\section{Introduction}

Cluster state quantum computation\cite{1wqc,mbqc} has become recently an attractive alternative to the standard quantum network model\cite{deutsch2,nc}, especially in the context of optical quantum computing.\cite{nielsen2,browne,benjamin,benjamin2,kieling,bk,beige,kok,louis} There are several advantages of using photonic qubits, including low decoherence, free-space propagation, availability of efficient single qubit gates and the prospect of miniaturization using optical silicon circuits.\cite{si_circuit} Cluster states with four\cite{photonic_1wqc} and six photons\cite{G6_cluster} have been experimentally prepared and characterized. Recently an 8-qubit photonic cluster state has been demonstrated in the context of topological quantum error correction.\cite{top_qec}

In order to be useful in quantum algorithms, we need to scale up these promising results to clusters containing tens to hundreds of encoded qubits. One of the present roadblocks towards this goal is the probabilistic nature (which implies postselection) of all the above schemes. Although linear optics schemes (like the KLM model\cite{klm,kok}) are in theory scalable and universal, they require single photon sources, photon number discriminating detectors, post-selection, fast feed-forward and quantum memories -- all these put severe experimental constraints. A possible solution to this problem is to have a deterministic architecture which avoids most of the aforementioned issues and can be easily scaled up.

The photonic module concept\cite{ph_module,ph_module2,ph_module3,hpqc} has been successful in showing how large cluster states can be prepared deterministically using a standard building block -- an atom in a cavity -- and classical switching. The atom in the cavity plays the role of an ancilla and provides the strong interaction required to couple the photons (the computational qubits). At this stage it is important to explore several designs in order to quantify resource requirements. Indeed, each particular architecture will involve complex trade-offs between design simplicity, total number of elementary operations and their accuracy plus other technological constraints (fabrication methods, operating environment etc).

Motivated by these considerations, in this article we explore an alternative architecture for constructing photonic cluster states with photonic modules. The original photonic module functions as a parity gate -- given $n$ photons as input, it performs a nondestructive parity measurement on the arbitrary photonic state.\cite{ph_module,ph_module2} This operation determines the blueprint of the optical circuit in terms of the number of layers and connectivity of basic building blocks, switching sequence, rerouting etc. In this article we examine an alternative photonic module build around the controlled-$Z$ gate $C(Z)$ instead of the parity gate and see how the design changes with this choice.

The structure of the article is as follows. In Section 2 we begin by discussing the two main approaches for building cluster states, using either stabilizer/parity measurements or controlled-$Z$ operations. These two paths lead to different photonic modules which we will call, respectively, the parity module and the $CZ$ module. In Section 3 we explore a new network design of a photonic circuit build around the $CZ$ module and we show how changing the fundamental entangling gate leads to a simplified circuit design for preparing a 2D cluster state. In Section 4 we introduce a passive switching mechanism and the corresponding network design. In Section 5 we describe how our scheme can be generalized to construct a 3D topological cluster state.

\section{Cluster states and photonic modules: two approaches}

At the core of the photonic module is the interaction between a photon and an atom in a cavity. In the model we are considering here the photons play the role of the computational qubits and the atom in the cavity serves as an ancilla mediating the coupling between the photons. As in the original photonic module concept, we assume the photon-atom interaction to perform a $C(Z)$ gate between the photonic (computational) and atomic (ancillary) degrees of freedom.\cite{CZ,ph_module,ph_module2} This gate is then sufficient to entangle the photonic qubits, as we will discuss in the following.

There are two ways of describing a cluster state and each description provides a different way of preparing the state in the lab. First, we can view the cluster state as a stabilizer state, hence we can prepare it by measuring $n$ stabilizer operators, one for each qubit/vertex. The stabilizer operator of vertex $i$ is $X_i \prod_{j\in neigh(i)} Z_j$, where $X_i, Z_j$ are the Pauli operators of a vertex and the product is over all nearest neighbours; thus for a 2D cluster state each stabilizer involves at most five photons. This is the approach taken in Refs.~\refcite{ph_module2,ph_module3} where a cluster state (two- or three-dimensional) is prepared by sending $n$ unentangled photons through an array of parity modules (or $P$-modules). A $P$-module consists of a cavity with an atom in the center and performs a nondestructive parity measurement on the photons, i.e., it projects the initial photonic state onto even (odd) parity states. To prepare a 2D cluster state each photon has to pass through five cavities\cite{ph_module2} (for a 3D cluster state this number is four\cite{ph_module3}). The architecture of the full circuit is rather complex, consisting of several layers of photonic modules and routing switches directing the photons in and out of the cavities.

In the second description the cluster state is prepared in two steps\cite{1wqc}: (a) all qubits are initialized in the state $\ket{+}^{\otimes n}$, with $\ket{+}= (\ket{0}+\ket{1})/\sqrt{2}$; (b) a controlled-$Z$ operation $C(Z)= \mbox{diag}(1,1, 1, -1)$ is applied to each pair of qubits sharing a link in the underlying graph $G$: $\prod_{(i,j) \in edges(G)} C(Z)_{ij} \ket{+}^{\otimes n}$.

\begin{figure}[t]
\centerline{\psfig{file=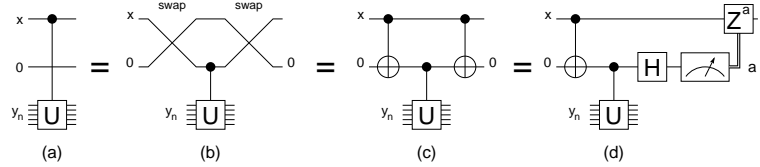, width=10 cm}}
\vspace*{8pt}
\caption{(a)-(d): equivalent quantum networks for a controlled gate $C(U)$ acting on $n$ qubits $y_n$. Since the ancilla starts in the $\ket{0}$ state, the SWAP gates in (b) are reduced to a pair of CNOT gates as in (c). In (d), the second CNOT in (c) (disentangling the ancilla) can be replaced by a measurement of the ancilla in the Fourier basis followed by a postprocessing gate $Z^a$ on the first qubit.}
\label{cz}
\end{figure}

This puts into perspective the difference between the two approaches -- in the first one the central resource is the parity gate, whereas in the second the $C(Z)$ gate. For photons measuring parity is in general easier than performing a $C(Z)$ gate. As photons do not interact directly, the usual way to perform a {\em deterministic} controlled-$U$ gate $C(U)$ between the two photons is to use an ancilla (e.g., an atom in a cavity) coupled to both, as in Fig.~\ref{cz}. The well-known solution is to first swap the first qubit and the ancilla, perform the $C(U)$ gate between the ancilla and the second qubit, and then swap back the ancilla and the first qubit; if the ancilla is prepared in the $\ket{0}$ state, this sequence requires only two CNOT gates and one $C(U)$ gate, as in Fig.~\ref{cz} (a)-(c). This procedure has been used to entangle two photons (the qubits) using an atom in a cavity (the ancilla).\cite{duan_kimble} The problem with this scheme is that the first photon has to interact twice with the cavity, first to entangle and subsequently to disentangle it from the ancilla, Fig.~\ref{cz}(c). This requires a photonic buffer to store the first photon until the appropriate time and then redirect it to the cavity, increasing the complexity. For this the reason the parity module was prefered as the central building block in previous schemes for constructing 2D\cite{ph_module2} and 3D photonic cluster states.\cite{ph_module3}

In this article we focus on the second approach of preparing a cluster state and use the $C(Z)$ gate as the main resource -- we call this the $CZ$ module. The first step is to notice that the second CNOT gate in Fig.~\ref{cz}(c) is not necessary, and that we can disentangle the first photon and the ancilla by measuring the ancilla in the $\{ \ket{+},\ket{-} \}$ basis, Fig.~\ref{cz}(d). Let's see how the quantum network in Fig.~\ref{cz}(d) works. After the first two gates, CNOT and $C(U)$, the initial state is transformed to $\ket{x0y_n}\rightarrow \ket{xxy_n} \rightarrow \ket{xx} U^x\ket{y_n}$. In order to disentangle the ancilla from the control qubit, we apply a Hadamard $H$ and then measure the ancilla; the previous state is first transformed to $\ket{x} (\ket{0}+ (-1)^x\ket{1}) U^x\ket{y_n}$ (after $H$) and then to $(-1)^{ax} \ket{x} \ket{a} U^x\ket{y_n}$ (after measurement, assuming the result is $a$). The extra phase is then removed by applying to the first qubit a feed-forward $Z^a$, such that the network in Fig.~\ref{cz}(d) performs the following transformation
\begin{equation}
\ket{x0y_n} \rightarrow \ket{x} \ket{a} U^x\ket{y_n}
\end{equation}
thus proving the circuit to be equivalent to a $C(U)$ between $x$ and $y_n$. Note that $y_n$ is an arbitrary state of $n$ qubits/qudits. Since $Z$ and $C(Z)$ gates commute, we can apply the corrective $Z^a$ action at the very end of the cluster state preparation, further simplifying the network.

\begin{figure}[t]
\centerline{\psfig{file=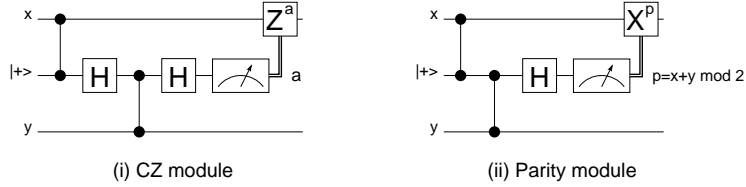, width=10 cm}}
\vspace*{8pt}
\caption{Two types of photonic modules implementing: (i) a $C(Z)$ gate, (ii) a parity gate. The qubits $x,y$ are photons, each interacting with an atom in a cavity ancilla (middle) initialized in the $\ket{+}$ state. A postprocessing gate is applied to the first qubit depending on the result of the measurement. In the case of the $CZ$ module the corrective $Z^a$ gate can be applied at the end of the cluster state preparation since $Z$ commutes with subsequent $C(Z)$ gates.}
\label{compare}
\end{figure}

Fig.~\ref{compare} shows the difference between the $CZ$ module and the parity module, as discussed above. The difference between the two is minimal -- only a Hadamard gate $H$ on the ancilla after the first qubit interaction. However, this minimal modification leads to a simplified circuit implementing a cluster state, as we will describe next.

\section{Building a 2D cluster: circuit design}

In this section we show how to use the $CZ$ module described above to build a 2D square lattice cluster state. As this state is a universal resource for quantum computation one can used it to perform an arbitrary quantum algorithm.

\begin{figure}[t]
\centerline{\psfig{file=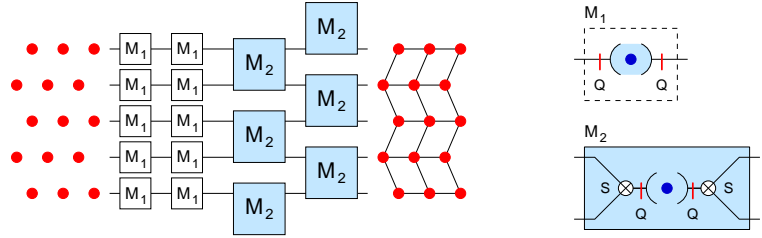, width=10 cm}}
\vspace*{8pt}
\caption{Left: Building a 2D photonic cluster state. Photons (red dots) enter from the left, prepared in the $\ket{+}$ state. Each photons passes through two $M_1$ and two $M_2$ modules. The $M_1 (M_2)$ applies a $C(Z)$ gate between a photon and its left/right (top/bottom) neighbours; these are indicated by black lines in the final cluster state. Right: The modules can be implemented as $Q$-switched cavities. The $M_2$ module has also two active switches $S$ which redirect the photons to the central cavity area and then back to their rails after interaction. The two switches $S$ are synchronous and can be controlled by a single flip-flop circuit.}
\label{2d_active}
\end{figure}

\begin{figure}[b]
\centerline{\psfig{file=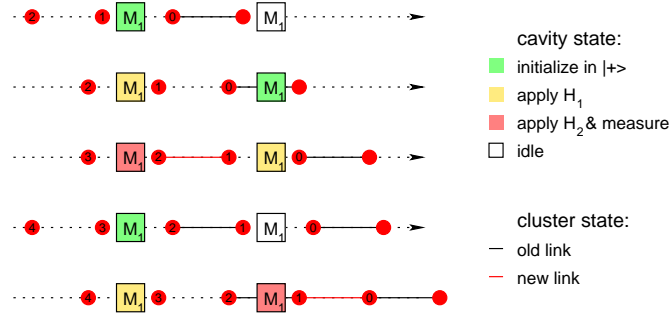, width=9 cm}}
\vspace*{8pt}
\caption{Time sequence of the action performed by two $M_1$ modules. The first (second) module applies a $C(Z)$ gate between photon pairs $(2k-1, 2k)$ and $(2k, 2k+1)$, respectively. The result is a linear cluster state which is then passed to $M_2$ modules to complete the 2D structure by adding the vertical links between qubits. $H_{1,2}$ are Hadamard gates, see Fig.\ref{compare}(i).}
\label{m1}
\end{figure}

Each node (qubit) in a square lattice has four neighbours so we need to apply four $C(Z)$ gates to each photon. The circuit architecture is shown in Fig.~\ref{2d_active}. Photons are prepared in the $\ket{+}$ state and pass through two $M_1$ and two $M_2$ modules. The temporal delay between photons on the same line is $T$; photons on adjacent lines of the cluster are delayed by half period $T/2$ in order to avoid them arriving simultaneously at the $M_2$ interaction region. We assume that $T$ is large enough such that the full cycle of operations of the cavity atom, between initialization and measurement, is contained within $T$. Each module contains an atom in a cavity. The $M_1$ modules act on the same line and apply a $C(Z)$ gate between a given photon and its left and right neighbours. In Fig.~\ref{m1} we show a time sequence of this action. The $M_2$ modules perform the same function between photons on different lines, hence they contain two switches $S$ to direct the photons from, and respectively back to, their rails before and after interacting with the cavity. The two switches $S$ are synchronous (there are both up or down at the same time) and can be controlled by a single flip-flop circuit.

Let's see now what are the resource requirements to prepare a $m\times n$ cluster state, with $n$ the horizontal dimension of the cluster, equal to the number of time steps. For each horizontal line we need two $M_1$ and one $M_2$ modules, hence the total number of $M_1$ and $M_2$ modules is, respectively, $2m$ and $m-1$ (the -1 comes from boundary effects). Each edge in the cluster involves a measurement of the atom in the cavity, hence the total number of measurements is $m(n-1)+n(m-1)=2mn-m-n$.

\section{Passive switching}

As we discussed before, the network in Fig.~\ref{2d_active} uses active switching in the $M_2$ modules. Although this can be implemented by a classical flip-flop (since both switches are synchronous), it requires external pulses and additional wiring, adding an extra layer of complexity. In this section we show how this classical routing can be eliminated completely by using {\em passive switching}.

Suppose the photons have an additional, non-computational, degree of freedom. We call it a 'tag' and for our purpose it is sufficient that it takes only two values. In order to have a passive switch, we need two conditions: (i) the atom-photon interaction is independent of and does not change this degree of freedom (i.e., it preserves the value of the tag) and (ii) there is a simple passive device which routes the photons according to their tag: say 0 on upper path and 1 on the lower path; equivalently, it reflects photons with tag 0 and transmits photons with tag 1. Conceptually, the network in Fig.~\ref{2d_active} can be described in the same framework: the photon time bin is the tag (remember photons on neighbouring rails are temporally offset) and the switch is the flip-flop, directing each photon according to their time stamp (but in this case the flip-flop does not qualify as 'passive').

The typical example we will use in the following is the polarization degree of freedom and a polarizing beam splitter (PBS): the PBS transmits $H$- and reflects $V$-polarized photons. Another example is a dichroic mirror which reflects, e.g., green photons and transmits red ones (however in this case it is difficult to engineer the atom-photon interaction such that the cavity is insensitive to photons' colour). A third example is orbital angular momentum and holographic plates which transmit/reflect photons according to their orbital angular momentum.\cite{oam} Of course, one can imagine various implementations of this scheme using other degrees of freedom.

\begin{figure}[]
\centerline{\psfig{file=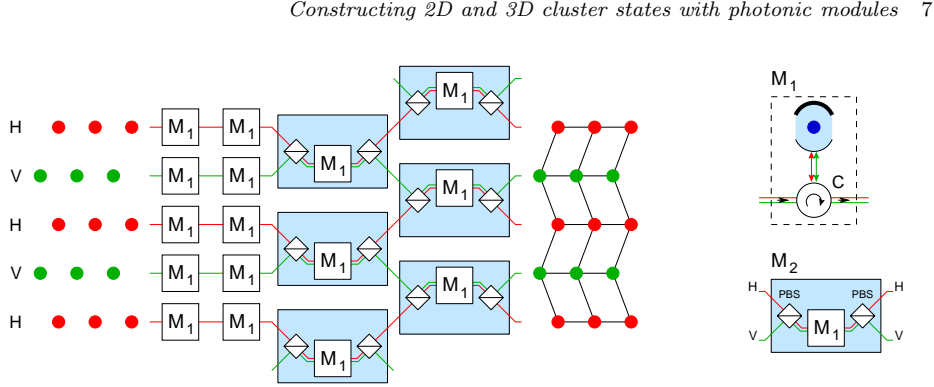, width=12 cm}}
\vspace*{8pt}
\caption{Left: Passive switching architecture for a 2D photonic cluster state. We show only the 1-rails of a dual rail encoding. Photons have an extra degree of freedom (tag) which is orthogonal for nearest-neighbours. If this tag is the polarization, photons are either $H$- or $V$- polarized (red and green, respectively). Assuming the atom-photon interaction is polarization preserving, switching in $M_2$ modules can be done passively with polarization beam splitters (PBS). Right: An alternative way of constructing the $M$ modules by reflecting a photon from a cavity using a circulator $C$; this construction, due to Duan and Kimble, eliminates $Q$-switching of the cavities.}
\label{2d_passive}
\end{figure}

In the following we use the polarization as a tag and the PBS as a passive switch. This requires two things. First, we need to use a mode (i.e., path encoded) qubit as the computational one. Second, the atom-photon interaction should be polarization preserving; an example of such interaction is $e^{i \theta (n_v+n_h) \sigma_z}$, with $n_v+n_h=n$ the total number of photons. 

Using path encoding qubits implies the photonic chip has to be placed between two beam-splitters in the 1-arm of a Mach-Zehnder interferometer. In Fig.~\ref{2d_passive} we show the new architecture of the chip -- note that only the 1-rails in the dual rail encoding are shown; the 0-rails are situated in a parallel plane of the chip. The operations are identical to the ones described before. The only difference is switching in the $M_2$ modules which is done passively by the PBSs. Since in this architecture all classical routing is done passively (i.e., without external control), the only control signals are build in $M_1$ modules, as $M_2$ modules are nothing but $M_1$ plus two switches, see Fig.~\ref{2d_passive}.

It is worth mentioning another feature of this design. The modules in the two (vertical) layers containing the $M_2$ modules are synchronized so the control signals for the cavities (initialization, Hadamard gates) can be done by a global control pulse applied to all $M_2$ modules. The only step in which individual control is still needed is the final readout of the cavities at the end of the $C(Z)$ gate (see Fig.~\ref{compare}(i)). One can still have a global readout pulse, provided each module has a local 1-bit memory which stores the measurement result of each cavity. With an appropriate design (we need to take into account the offset between green and red photons) one can envisage global control of {\em all} modules in the optical chip. In this case we can eliminate the individual control lines for each module; this becomes especially important in a 3D layout (see next section), when addressing a particular module buried inside the chip is difficult.

One can think of two different designs for implementing the $C(Z)$ gate between the atom and the photon. The first uses a $Q$-switched cavity\cite{ph_module,ph_module2}, as in Fig.~\ref{2d_active}: photons enter the cavity through the left, interact with the atom and then are $Q$-switched out to the right. The second employs the scheme of Duan and Kimble\cite{duan_kimble} -- photons are reflected from the cavity (the lower mirror is partially reflective) and exit through the same port (Fig.~\ref{2d_passive}, right). In this case an optical circulator redirects the photons to the exit rail. As a consequence, in this variant of the design we eliminate both the classical routing and the $Q$-switching of the cavities.

\section{Preparing a 3D topological cluster state}

Topological cluster state computation pioneered by Raussendorf, Harrington and Goyal\cite{top_cluster,top_cluster2} has attracted recently considerable interest as a fault-tolerant architecture for constructing photonic cluster states.\cite{ph_module3} Here we briefly discuss how to adapt the previous 2D network design to a 3D setup.

\begin{figure}[t]
\centerline{\psfig{file=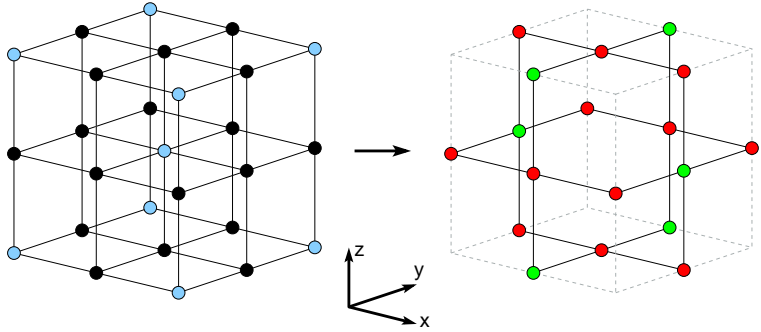, width=10 cm}}
\vspace*{8pt}
\caption{Starting with a regular cubic lattice (left), we construct the 3D topological code (right) by removing the blue photons together with all their links.}
\label{3d}
\end{figure}

The 3D topological cluster state\cite{top_cluster,top_cluster2} can be constructed from a regular cubic lattice (Fig.~\ref{3d}, left) by removing the blue qubits together with all their links, as in Fig.~\ref{3d}, right. One way of doing this is to construct first a 3D cubic cluster state and subsequently measure the blue qubits in the $S_z$ basis. The measurement eliminates the blue photons and their links from the lattice, but is not efficient since it involves feed-forward (we have to take into account the result of the measurement).

\begin{figure}[t]
\centerline{\psfig{file=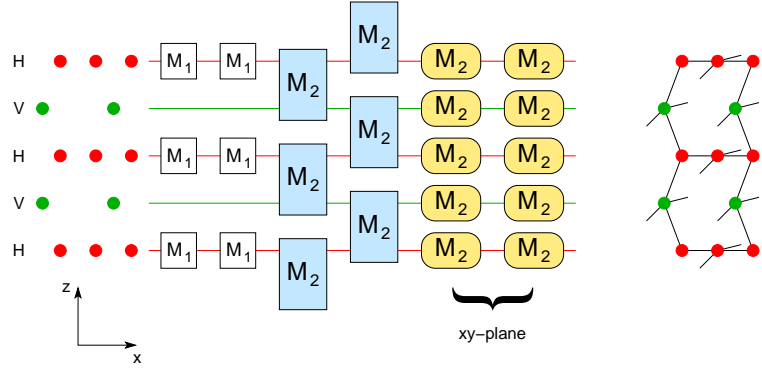, width=10 cm}}
\vspace*{8pt}
\caption{Left: network for a 3D topological cluster state, viewed in the $xz$-plane; photons flow along $x$ axis. The yellow $M_2$ modules are oriented in the $xy$ plane and couple two adjacent photons in the $y$ direction. The green photons are doubly spaced compared to the red ones; note there are no $M_1$ modules on the green lines. Right: schematics of a 3D topological cluster state; each photon has only 4 links.}
\label{3d_module}
\end{figure}

A better way of obtaining the same result is the following. Suppose we have photonic module which prepares a regular cubic cluster state. We can extend the 2D module described in the last section in a straightforward way to a 3D geometry by adding two extra layers of $M_2$ modules in order to couple each photon to its nearest neighbours along the third spatial direction, orthogonal to the $xz$-plane, as in Fig.~\ref{3d_module}. However, instead of eliminating the blue photons by measurement after the photonic modules, we remove the blue photons at the injection stage, i.e., we run the photonic module with some of the photons missing. Thus the photon sources injecting the green photons run at half the frequency of the sources injecting the red photons, as in Fig.~\ref{3d_module}. Moreover, since the green photons don't have links with other green photons, we can eliminate completely the $M_1$ modules on the green lines. Each green photon will pass through four $M_2$ modules, two for the links in the $xz$-plan and the other two for the links in the $xy$-plan. The red photons pass thorough six modules, two for each spatial direction, in a straightforward generalization of the 2D case. However, since now half of the green photons are missing, each red photon will have only four links: two links in the $x$ direction (always) and another two with the green photons, either along $y$ or $z$ axis. The resulting state (Fig.~\ref{3d}, right) is exactly the 3D topological cluster state from Ref.~\refcite{ph_module3}. Once prepared, the 3D topological cluster state can be used as a universal, fault-tolerant resource for QIP.

\section{Conclusions}

In this article we described a scheme for preparing large scale photonic cluster states with photonic modules. In our model we implement directly a $C(Z)$ gate between two photons using as an ancilla an atom in a cavity. Compared to the original photonic module design which uses a parity gate\cite{ph_module}, this choice of entangling gate leads to a simplified architecture with fewer modules ($3m-1$ compared to $5m$ for a 2D cluster, with $m$ the width of the cluster) and classical switching. Moreover, if the atom-photon interaction is polarization preserving there is no need for active switching at all. In this case one can have only passive switching, e.g., using polarising beam splitters and photons in neighbouring rails having orthogonal (H/V) polarization. This passive switching completely eliminates the need of an active switching mechanism synchronized with the photons, thus reducing the complexity and the associated decoherence. Another feature of the present design is the possibility of using global control of {\em all} modules in the optical chip. This becomes especially important in a 3D layout, when addressing a particular module buried inside the chip is difficult.

The model discussed here paves the way towards integrated photonic circuits\cite{si_circuit} on a chip as a basis for future quantum optical processors. Even with a small to medium number of photonic qubits available, such a chip will be useful as a quantum repeater\cite{dlcz,repeater,repeater2} or as an element in a future quantum internet\cite{q_internet} architecture.

\section*{Acknowledgments}

We acknowledge financial support from EU projects QAP and HIP and Japan MEXT.


\end{document}